\documentclass[10pt,a4paper]{article}
\usepackage{graphicx}
\usepackage{float}
\usepackage{amsmath}
\usepackage{amssymb}
\usepackage{bm}
\usepackage{a4wide}
\usepackage[affil-it,auth-sc]{authblk}
\usepackage{times}

\newcommand{\rmd}{{\rm d}}

\title{The quantum Hall effect under the influence of a top-gate and integrating AC lock-in measurements}

\author[1,2]{Tobias Kramer}
\author[2,3]{Eric J.\ Heller}
\author[4]{Robert E.\ Parrott}

\affil[1]{Institute for Theoretical Physics, University of Regensburg, 93040 Regensburg, Germany}
\affil[2]{Department of Physics, Harvard University, Cambridge, MA~02138, USA}
\affil[3]{Department of Chemistry and Chemical Biology, Harvard University, Cambridge, MA~02138, USA}
\affil[4]{School of Engineering and Applied Science, Harvard University, Cambridge, MA~02138, USA}

\date{May 20, 2009}

\begin{document}

\maketitle

\begin{abstract}
Low frequency AC-measurements are commonly used to determine the voltage and currents through mesoscopic devices. We calculate the effect of the alternating Hall voltage on the recorded time-averaged voltage in the presence of a top-gate covering a large part of the device. The gate is kept on a constant voltage, while the Hall voltage is recorded using an integrating alternating-current lock-in technique. The resulting Hall curves show inflection points at the arithmetic mean between two integer plateaus, which are not necessarily related to the distribution of the density of states within a Landau level.
\end{abstract}

\section{Introduction}

In Ref.~\cite{Kramer2008c}, we reported evidence for inflection points of the Hall resistivity at half-integer filling factors, where the slope of $\partial \rho_{xy}/\partial B$ goes through a local minimum. We attributed the existence of these inflection points to features of the LDOS in the presence of a strong potential gradient near the injection corner of the device. In march 2009, new experiments in a different sample have shown additional features, which require a new interpretation of the earlier data as well as of the new experimental results. In particular, after modelling the measurement protocol in every detail, we propose a different interpretation of the inflection points shown in Ref.~\cite{Kramer2008c}. For an alternating current flow we predict oscillatory shifts of the Fermi energy in the two-dimensional electron system due to the presence of a non-zero Hall voltage. The oscillatory shifts are quantitatively calculated in this manuscript and provide a different explanation of the data shown in Ref.~\cite{Kramer2008c}. We do not rule out the existence of inflection points caused by the LDOS in direct-current measurements.

\section{AC measurement protocol}

In order to obtain a better signal-to-noise ratio, alternating-current (AC) measurements are often preferred to direct-current (DC) measurements. While there are different ways to record and process the AC signal, one commonly used protocol is to use a $\cos(\omega t)$-AC wave-form for the current between source and drain of a Hall device 
\begin{equation}
I(t)=\hat{I}\cos(\omega t)=\sqrt{2}I_{rms}\cos(\omega t),
\end{equation}
and to record the Hall voltage by integrating the instantaneous Hall voltage over one period of the cosine signal. In order to amplify only signals matching the frequency of the current-modulation, the oscillating $\cos(\omega t)$-signal is fed into the integration loop. Thus the read-out Hall voltage of the AC-lock-in amplifier becomes
\begin{equation}\label{eq:lockin}
V^{\text{read}}_{H}=c \int_0^{2\pi/\omega}\rmd t\; R_{xy}[I(t)]\;I(t)\;\cos(\omega t),
\end{equation}
where $c$ denotes the normalization, determined experimentally by comparison with a known resistance.

\section{Theory of the AC measurement in a sample with top-gate}

\begin{figure}[t]
\begin{center}
\includegraphics[width=0.8\textwidth]{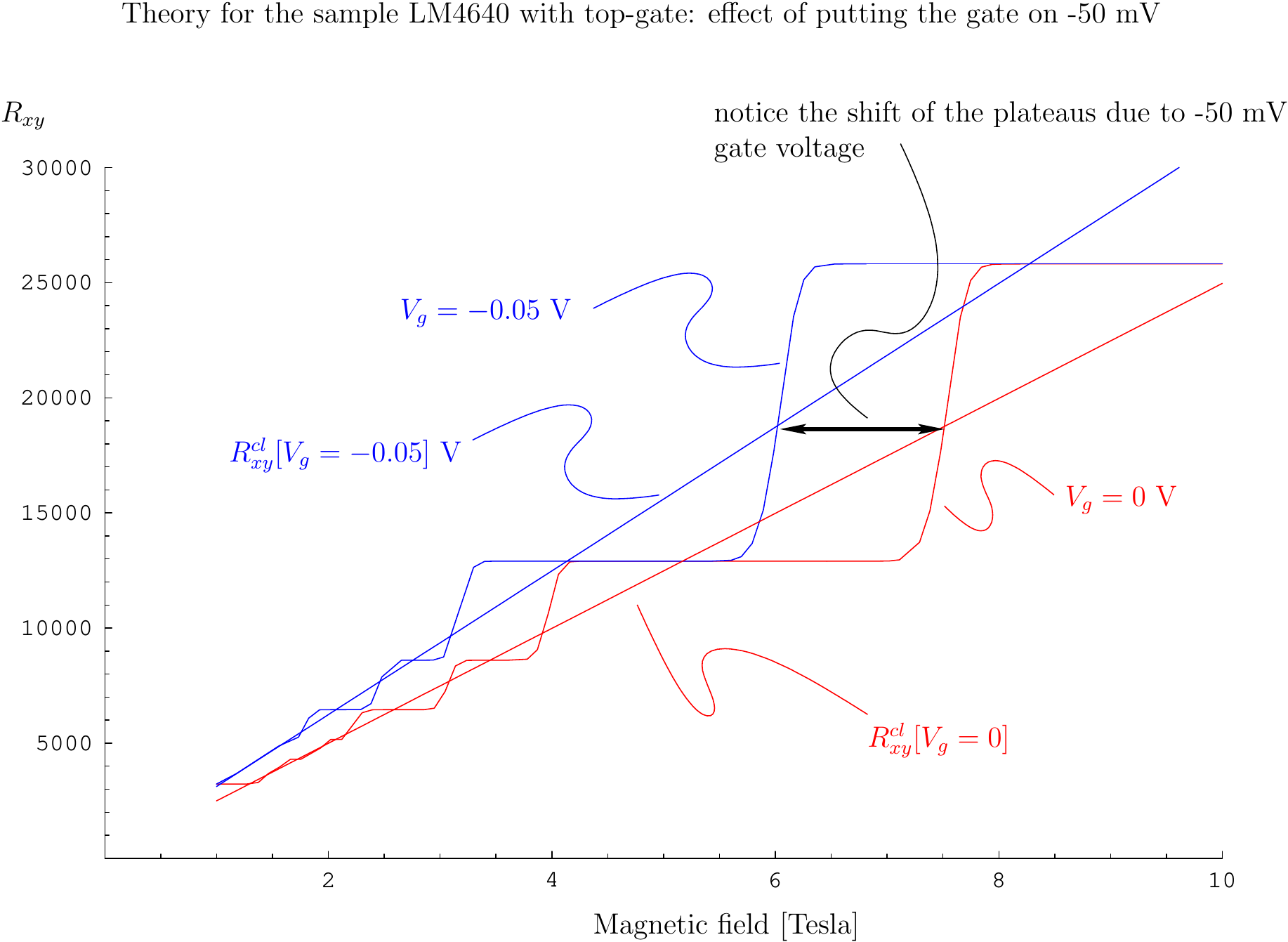}
\end{center}
\caption{Effect of applying a fixed voltage to the top-gate (theory). A negative gate-voltage reduces the density and leads to a steeper Hall curve.}
\label{fig:rxyg}
\end{figure}

The top-gate is kept on a constant potential (the same potential as one of the Ohmic contacts), while the voltage of the second Ohmic contact is oscillating. This leads to an oscillating AC current through the device. In the presence of a magnetic field the Hall voltage forms across the sample in response to the current. The Hall voltage is also oscillating in phase with the current.
The maximum Hall voltage during the oscillation period is given by
\begin{equation}
\hat{V}_H=\hat{I} R_{xy}({\cal B}),
\end{equation}
and yields for $I_{rms}=1\;\mu$A at $R_{xy}=19$~k$\Omega$ a peak Hall voltage of $\hat{V}_H=27$~mV. The range of variation of the Hall voltage is thus $2\times27$~mV$=54$~mV. The measured sample LM4640 has been grown with the following layer sequence:\\[1.5ex]
\begin{center}
\begin{tabular}{l|l}
Material & Thickness \\\hline
Au/Ti gate & on top \\
Undoped GaAs cap & $10$~nm \\
Si doped Al$_{0.33}$Ga$_{0.67}$As ($N_{\text{Si}}=2.65\times 10^{24}$~m$^{-3}$) & $40$~nm \\
Undoped spacer Al$_{0.33}$Ga$_{0.67}$As & $20$~nm \\
Undoped GaAs & $1\;\mu$m
\end{tabular}
\end{center}
\begin{figure}
\begin{center}
\includegraphics[width=0.499\textwidth]{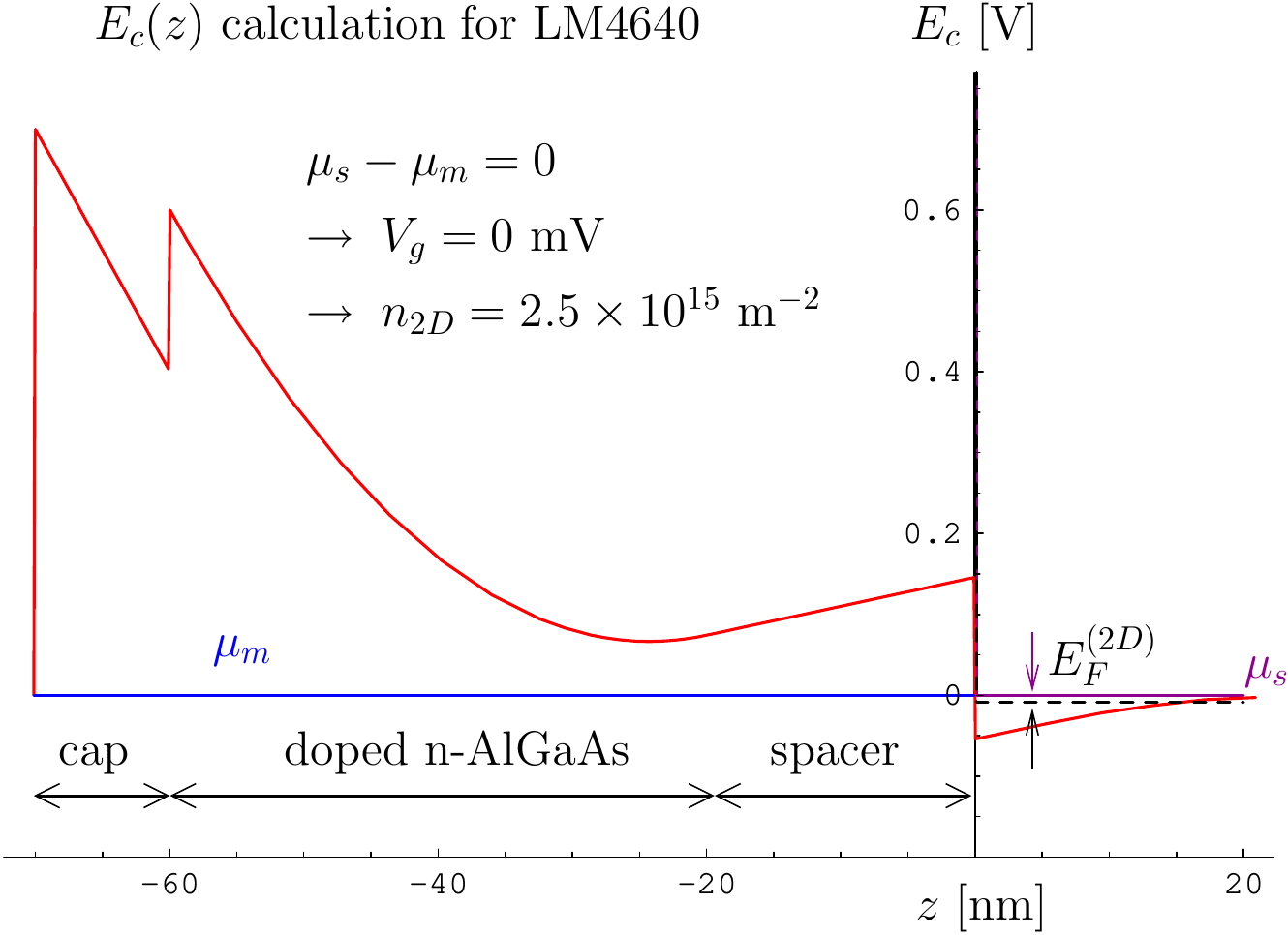}\hfill
\includegraphics[width=0.499\textwidth]{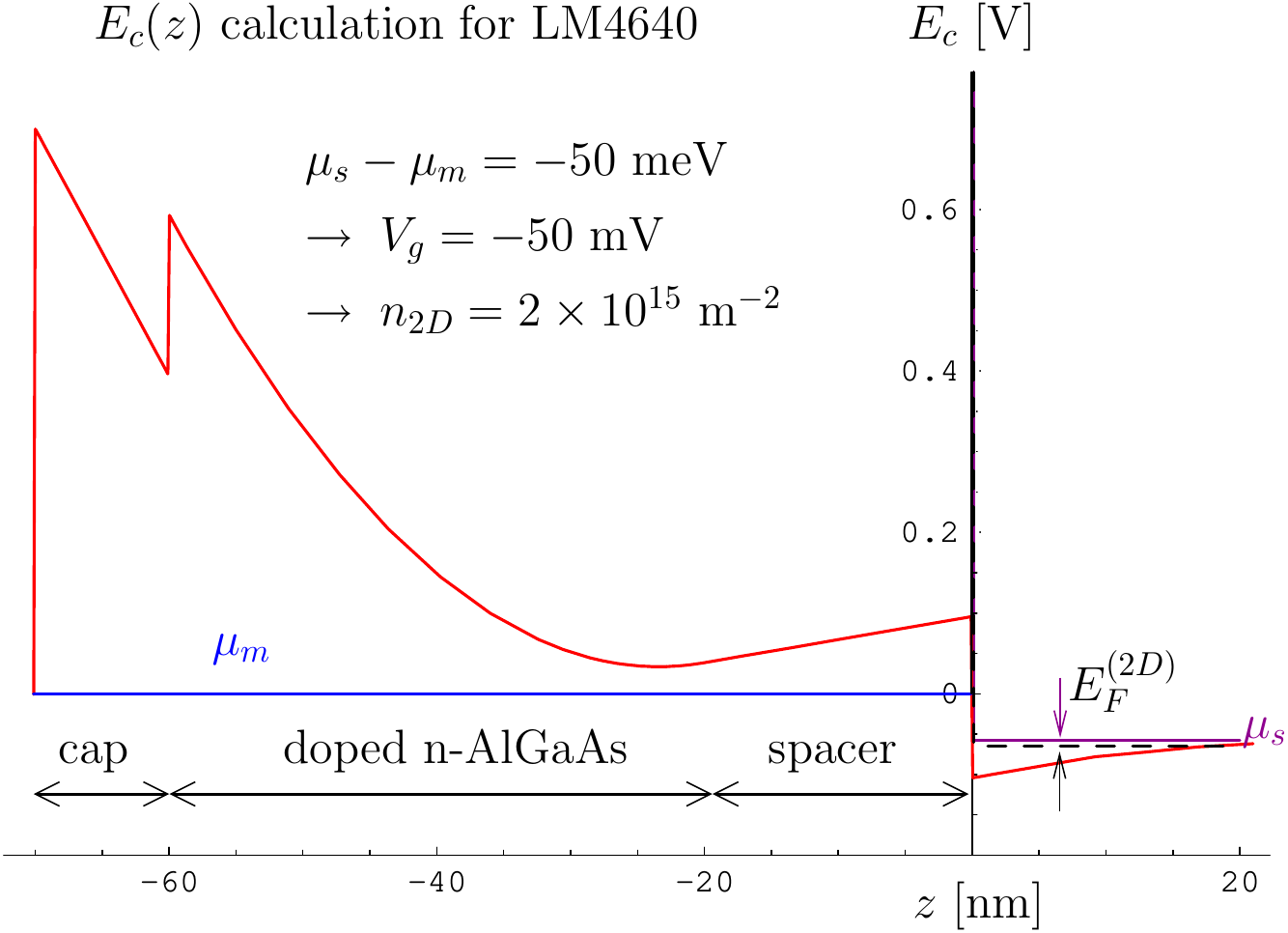}
\end{center}
\caption{Calculated conduction band profile $E_c(z)$ along the growth $z$-direction of the sample LM4640. 
Left panel: In the absence of a potential difference between gate potential $\mu_m$ and 2DEG potential $\mu_s$, the density is $n^{2D}=2.5\times 10^{15}$~m$^{-2}$ and $E_F^{2D}=8.9$~meV. 
Right panel: A potential difference of $50$~meV between $\mu_m$ and $\mu_s$ (by putting a negative voltage of $-50$~mV on the top-gate) shifts the quantum well and reduces the density to $n^{2D}=2.0\times 10^{15}$~m$^{-2}$, while the Fermi energy shifts to $E_F^{2D}=7.1$~meV.
}\label{fig:ecvg}
\end{figure}
~\\
We have calculated, based on \cite{Davies1998a} eq.~(9.6), the conduction-band structure of the sample using the triangular well approximation for the zero-point energy of the quantum well and under the assumption that a fraction of $22/100$ Si atoms gets ionized. Part of the electrons will contribute to the 2DEG, while another part will contribute to the surface states. We assume that the surface states lead to an offset of $0.7$~V. The resulting conduction band diagram for zero gate voltage $V_g$ and for a negative gate voltage of $V_g=-50$~mV is shown in Fig.~\ref{fig:ecvg}. Note that the gate voltage is \textit{defined} as the difference in potentials of the metallic top-gate $\mu_m$ and the potential of the two-dimensional electron gas (2DEG) $\mu_s$:
\begin{equation}\label{eq:vg}
V_g:=\frac{\mu_m-\mu_s}{e}.
\end{equation}
The calculated values agree very well with the measured densities inferred from Hall experiments at gate voltages $0$ and $-50$~mV, shown in Fig~\ref{fig:rxyg}. The possibility to control the density of the 2DEG by applying a gate-voltage and shifting $\mu_m$ is well-known.

Interestingly, eq.~(\ref{eq:vg}) shows that the ``gate-voltage'' can be set by either changing $\mu_m$ or changing $\mu_s$. A change in $\mu_s$ does not usually occur in a 2DEG, but the Hall effect is an important exception from this rule. In a Hall device, the 2DEG is not on a single potential across the device, rather the potential of the 2DEG drops from one side to the other side of the device by the Hall potential $\mu_H$. Thus the presence of the Hall potential generates a non-zero difference between the metallic gate at potential $\mu_m$ and the 2DEG at potential $\mu_s+\mu_H$. The Hall potential increases for fixed magnetic field with increasing current. From eq.~(\ref{eq:vg}) we deduce that a change of the potential of the 2DEG is equivalent to applying a gate voltage and thus adjusting the density and the average slope of $R_{xy}({\cal B})$.

The last observation bears important consequences for the interpretation of AC-experiments at non-zero currents, which induce a possibly large Hall voltage: the recorded Hall voltage is a superposition of all Hall voltages present during one period of the integration. The AC-experiment in the presence of a top-gate, which is kept on a constant potential $\mu_m$, records the (weighted) mean of many Hall-curves, where each individual Hall curve is effectively measured at a different gate voltage, since the Hall potential oscillates in phase with the current: 
\begin{equation}
\mu_{\text{2DEG}}=\mu_s+\mu_H(t).
\end{equation}
\begin{figure}[t]
\begin{center}
\includegraphics[width=0.8\textwidth]{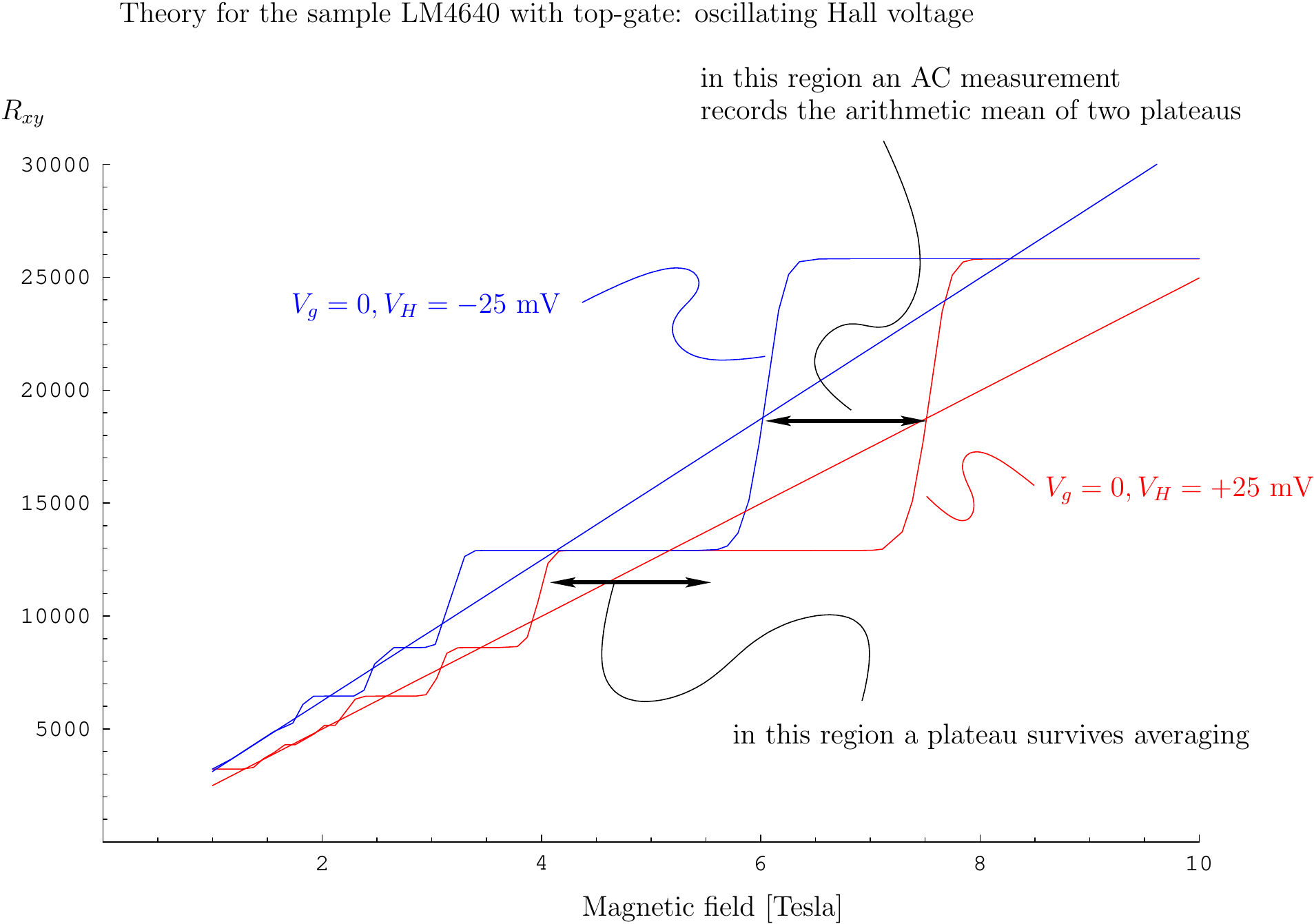}
\end{center}
\caption{Effect of the oscillating Hall voltage underneath top-gate kept on a constant potential (theory). Integrating over one period of the AC-lock-in amplitude records the (weighted) mean of different Hall curves. Note that the figure is schematic, since $V_H$ is dependent on the magnetic field and is thus the change in slope of the Hall curve is less pronounced at lower magnetic fields. The complete simulation, which takes the Hall-voltage dependence into account, is shown in the appendix.}
\label{fig:rxyo}
\end{figure}
The effect of the oscillations of $\mu_s+\mu_H(t)$ on the quantized Hall curves is precisely the same as putting an oscillatory voltage on the top-gate during the integration process of the lock-in measurement, see Fig.~\ref{fig:rxyo}. In Fig.~\ref{fig:rxyo} extended regions are visible, where the lock-in integration will lead to a read-out voltage corresponding to the arithmetic mean of two adjacent Hall plateaus (in LM4640 this occurs around ${\cal B}=6$~T, ${\cal B}=4$~T, and ${\cal B}=3$~T).

The implications for experiments are that inflection points are induced by the AC-protocol at values of 
\begin{equation}
R_{xy}^{\text{inflection}}=25812~\Omega\; \frac{1}{2}\; \left(\frac{1}{n}+\frac{1}{n+1}\right),\quad n=1,2,3,\ldots,
\end{equation}
corresponding to the arithmetic mean of two adjacent Hall plateaus. Converted to filling factors, we expect AC-inflection points at
\begin{eqnarray}
\nu_{12}&=&{\left[\frac{1}{2}\left(1+\frac{1}{2}\right)\right]}^{-1}=4/3\\
\nu_{23}&=&{\left[\frac{1}{2}\left(\frac{1}{2}+\frac{1}{3}\right)\right]}^{-1}=12/5\\
\nu_{34}&=&{\left[\frac{1}{2}\left(\frac{1}{3}+\frac{1}{4}\right)\right]}^{-1}=24/7
\end{eqnarray}
Indeed a close inspection of the experimental record shows that in an AC-measurement inflection points occur at the arithmetic mean
\begin{center}
\begin{tabular}[h]{ll}
$R_{xy}^{\text{inflection}}$ & Experiment $[\Omega]$ \\\hline 
$\nu_{12}$ at 19359~$\Omega$ & 19320~$\Omega$ ($\nu=1.34$)\\
$\nu_{23}$ at 10755~$\Omega$ & 10850~$\Omega$ ($\nu=2.38$)\\
$\nu_{34}$ at  7528~$\Omega$ &  7450~$\Omega$ ($\nu=3.46$)
\end{tabular}
\end{center}
The LDOS signature analyzed in Ref.~\cite{Kramer2008c} predicted inflection points at
\begin{equation}
R_{xy}^{\text{inflection, LDOS}}=25812~\Omega \frac{1}{n+1/2},\quad n=2,3,
\end{equation}
corresponding to the following filling factors
\begin{eqnarray}
\nu_1^{\text{LDOS}}&=&\frac{5}{2}\\
\nu_2^{\text{LDOS}}&=&\frac{7}{2}
\end{eqnarray}
and values of $R_{xy}$:
\begin{center}
\begin{tabular}[h]{ll}
$R_{xy}^{\text{inflection, LDOS}}$                       & Experiment \\\hline 
not expected (only intersection point at 17208~$\Omega$) & 19320~$\Omega$ \\
$\nu_1=5/2$ at 10325~$\Omega$                            & 10850~$\Omega$ \\
$\nu_2=7/2$ at  7375~$\Omega$                            &  7450~$\Omega$ 
\end{tabular}
\end{center}

\section{Conclusions}

The new experimental data and the detailed model of the AC-measurement protocol suggest a new interpretation of the data presented in Ref.~\cite{Kramer2008c}:
\begin{enumerate}
\item AC averaging explains an additionally observed feature at $\nu=4/3$, not contained in the LDOS theory.
\item The observed inflection points may not be the signature of the LDOS. The observed inflection points are 
closer to the values $\nu=4/3$, $\nu=12/5$, and $\nu=24/7$, predicted by the AC-model of the top-gate, than 
to the values $\nu=5/2$ and $\nu=7/2$, predicted by the LDOS theory.
\item The overall agreement of the AC averaging model with respect to the width of the Hall plateaus and the 
slope of the Hall curve is very good (see comparison in the Appendix).
\end{enumerate}

It is interesting to note that the gate keeps the two-dimensional Fermi-energy fixed, and no assumption of disorder is required to describe the physics of the device and to calculate the Hall curves shown in the appendix. Thus we do not rule out the possibility to construct a modified injection model of the QHE, which incorporates the current injection process and the Hall field in its foundations.  Further DC experiments are under discussion to reveal the shape of the LDOS in the injection region.

\providecommand{\url}[1]{#1}

\appendix

\section{Comparison of Hall curves}

On the next pages we compare the theoretical prediction of the LDOS theory of Ref.~\cite{Kramer2008c}, and the AC-averaging theory given in the present manuscript.

\pagebreak 
\begin{figure}[H]
\includegraphics[width=0.99\textwidth]{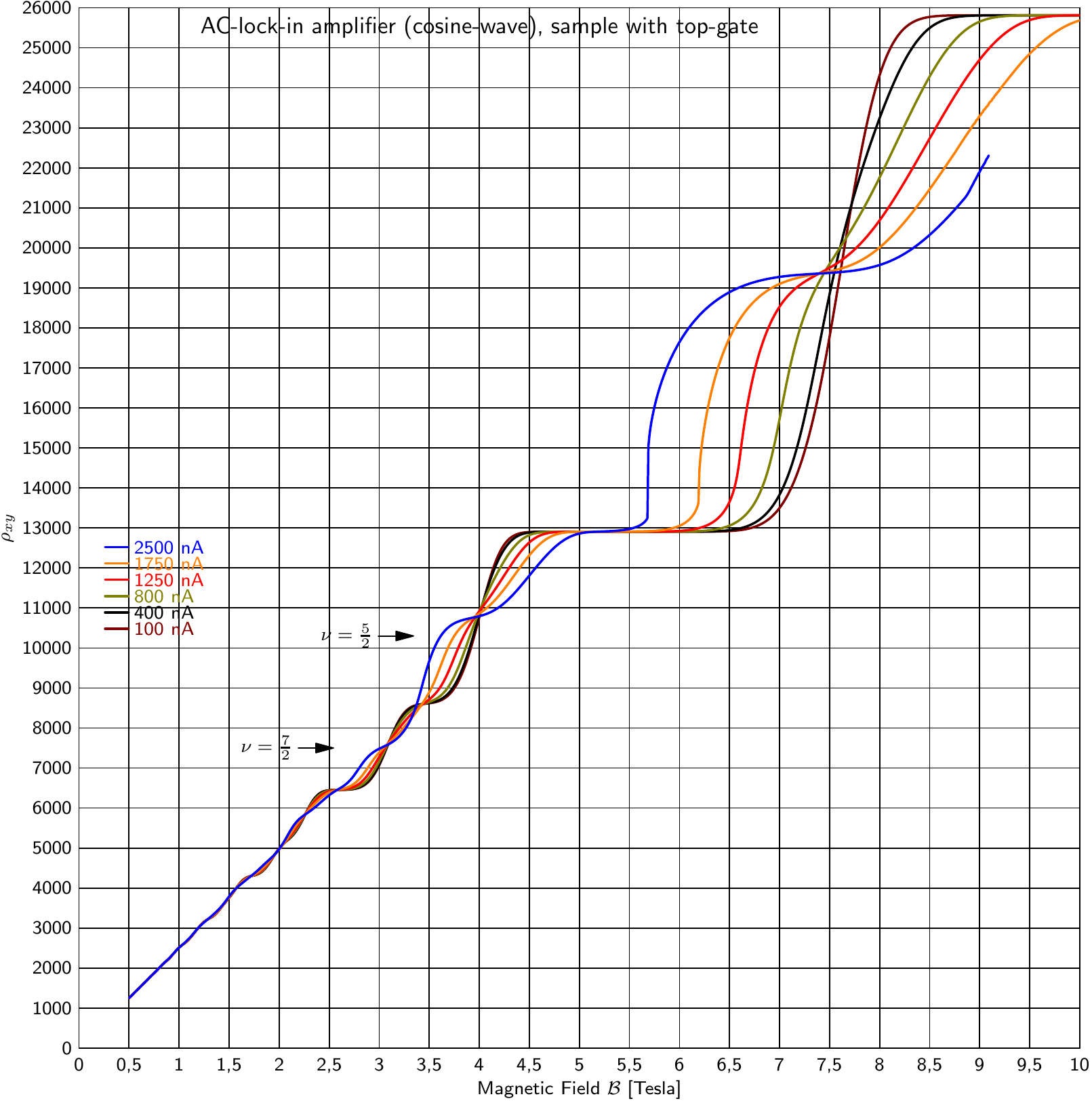}\\
\caption{Theory: AC-model with a featureless Gaussian broadened DOS. The legend denotes the value of $I_{rms}$. The calculation takes the change of the Hall voltage with magnetic field into account and the effective gate-voltage caused by the Hall potential, even though the gate is kept at a fixed potential. The curves are directly comparable to the experimentally recorded ratio $V^{\text{read}}_H/I_{rms}$ obtained with an AC-lock-in amplifier.}
\label{fig:rxynt}
\end{figure}

\pagebreak 
\begin{figure}[H]
\includegraphics[width=0.99\textwidth]{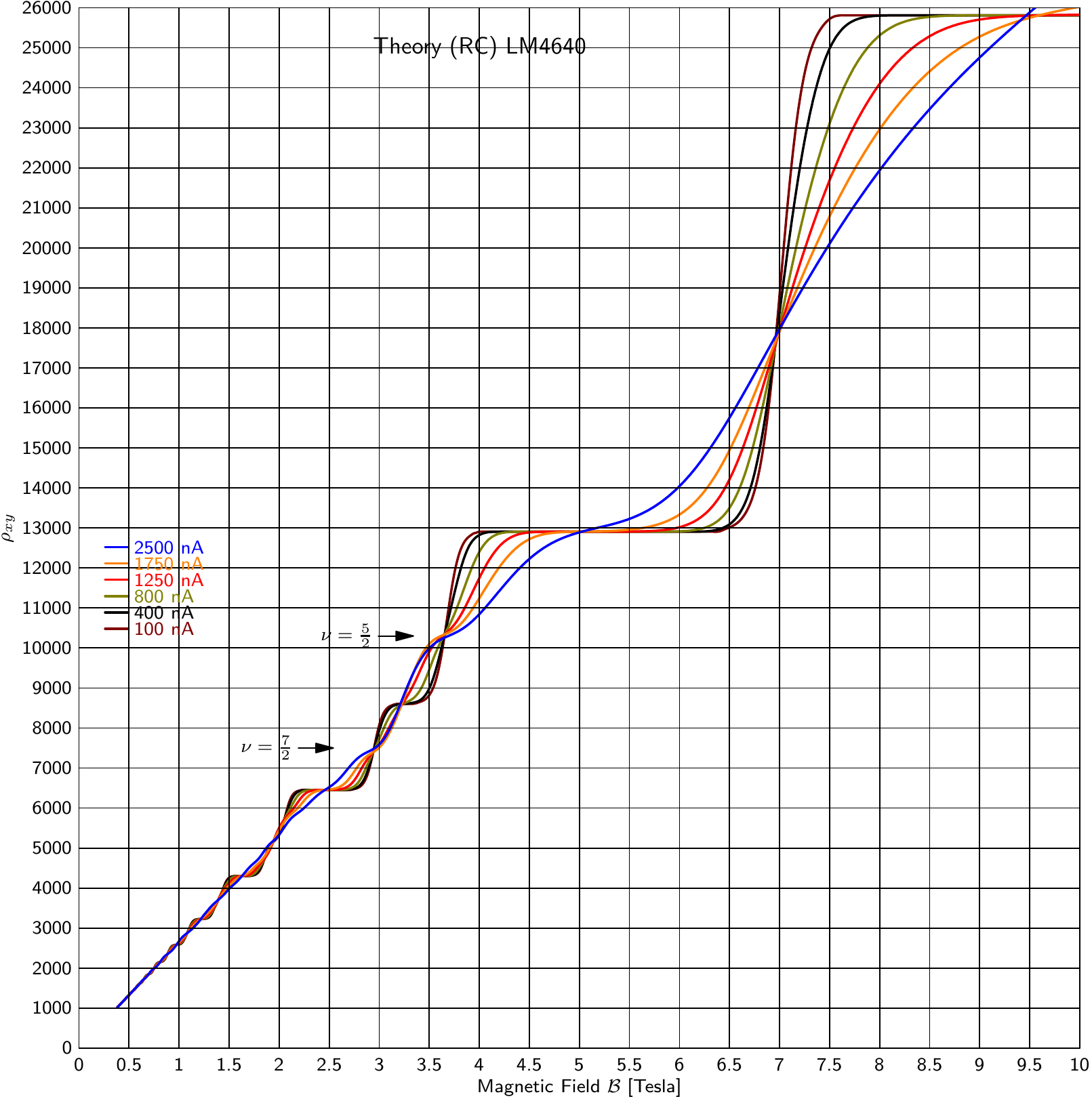}\\
\caption{Theory: model of Ref.~\cite{Kramer2008c}, LDOS modulation. Note that this is a DC calculation.}
\label{fig:rxyot}
\end{figure}

\end{document}